\documentclass{KapProc} 

\usepackage{epsf}
\usepackage{psfig}
\usepackage{graphicx}
\normallatexbib

\begin{document}

\articletitle{Kondo effect in non-equilibrium} 

\articlesubtitle{Theory of energy relaxation induced by
dynamical \\ defects in diffusive nanowires}

\author{Johann Kroha}
\affil{
Institut f\"ur Theorie der Kondensierten Materie and\\
Interfakultatives Institut f\"ur Nanotechnologie, Universit\"at 
Karlsruhe,\\ D-76128 Karlsruhe, Germany}
\email{kroha@tkm.physik.uni-karlsruhe.de}

\chaptitlerunninghead{Kondo effect in non-equilibrium }

\begin{abstract}
In diffusive Cu and Au quantum wires at finite transport voltage $U$ 
the non-equilibrium distribution function $f(E,U)$ exhibits  
scaling behavior, $f(E,U)=f(E/eU)$, indicating
anomalous energy relaxation processes in these wires. 
We show that in nonequilibrium 
the Kondo effect, generated either by magnetic impurities
(single-channel Kondo effect) or possibly by non-magnetic, degenerate 
two-level systems (two-channel Kondo effect), produces such scaling behavior 
as a consequence of a Korringa-like
(pseudo)spin relaxation rate $\propto U$ and  
of damped powerlaw behavior of the impurity spectral density as a
remnant of the Kondo strong coupling regime at low
temperature but high bias. The theoretical, scaled distribution 
functions coincide quantitatively with the experimental results, 
the impurity concentration being the only adjustable parameter.
This provides strong evidence for the presence of Kondo defects,
either single- or two-channel, in the experimental systems.
The relevance of these results for the problem of 
dephasing in mesoscopic wires is discussed briefly. 
\end{abstract}

\section{Introduction}
Recently, it has become possible to determine experimentally 
\cite{pothier.97,pothier.00} the distribution function 
of quasiparticles, $f_x(E,U)$, in dependence of the particle energy $E$
in a metallic, diffusive nanowire in stationary non-equilibrium 
with a finite transport voltage $U$ applied between the ends of the wire.
The measurements were done by attaching an additional superconducting
tunneling electrode 
at a position $x$ along the wire, so that $f_x(E,U)$ could be extracted 
from the tunneling current
\begin{eqnarray}
j_t=\frac{e}{h} \gamma N_o \int dE 
\Bigl[f_x(E,U) - f^o(E + eU_t )\Bigr] N_{BCS}(E+eU_t) \ .
\label{eq:current}
\end{eqnarray}
Here $N_{BCS}$ is the peaked BCS density of states in the superconducting
electrode, $N_o$ and $\gamma$ are the approximately constant density of 
states in the
wire and the tunneling matrix element, respectively, and 
$U_t$ denotes the voltage between the wire and the tunneling electrode.
$f^o(E) = 1/({\rm e}^{E/k_BT} + 1)$ is the Fermi distribution at 
temperature $T$.  

As expected for a diffusive wire with elastic impurity scattering 
\cite{nagaev.92},  $f_x(E,U)$ has a double-step shape, corresponding to 
the two chemical potentials at $E=0$ and $E=eU$ at the ends of the wire.
In addition, the steps are rounded, indicating
inelastic scattering processes in the wire. In Cu \cite{pothier.97}
and Au \cite{pothier.00} wires, the observed rounding is such that
the quasiparticle distribution obeys a remarkable scaling property
when $U$ exceeds a certain low-energy scale, 
$eU\stackrel{>}{\sim} eU_o\approx 0.1meV$:
$f_x(E,U) = f_x(E/eU)$ \cite{pothier.97}.
One may gain phenomenological insight into the nature of the
inelastic relaxation processes by observing \cite{pothier.97} that the scaling
property of $f_x(E,U)$ implies that the equation of motion of $f_x(E,U)$,
the quantum Boltzmann equation, and consequently the inelastic 
single-particle collision rate $1/\tau$ are scale invariant as well.
Assuming a (yet to be determined) two-particle potential 
$\tilde V(\varepsilon )$
with energy transfer $\varepsilon$,
$1/\tau$ is in 2nd order perturbation theory given as 
\begin{equation}
\frac{1}{\tau (E)} \equiv \frac{1}{\tau (E/eU)} \simeq
N_o^3 \int d\varepsilon  \int d\varepsilon '
|\tilde V (\varepsilon )|^2 F \Bigl(
\frac{E}{eU},\frac{\varepsilon}{eU},
\frac{\varepsilon '}{eU} \Bigr) \ ,
\label{perturb}
\end{equation}
where $F$ is a combination of distribution functions $f_x$ ensuring
that there is only scattering from an occupied into an unoccupied state.
The experimental finding that $f_x$ and hence $1/\tau$
depend only on the dimensionless
energies, as indicated in Eq.~(\ref{perturb}), implies that in the term
on the right-hand side only dimensionless energies (normalized to $eU$)
occur, i.e.~$\tilde V(\varepsilon ) \propto 1/\varepsilon $ in the 
scaling regime $|E| \stackrel{<}{\sim}k_BU_o\stackrel{<}{\sim}eU$.
It follows that within 2nd order perturbation theory (Eq.~(\ref{perturb})) 
the energy relaxation rate has a logarithmic energy dependence,
$1/\tau (E) \propto {\rm ln}(E/eU_o)$, non-vanishing at the Fermi energy.
This anomalous behavior has raised considerable interest in the
energy relaxation measurements \cite{pothier.97}, especially
because of the possible relation of the apparently
non-vanishing energy relaxation rate to the problem of dephasing 
saturation observed at low $T$ in the magnetoresistance of
nanoscopic wires \cite{mohanty.97,gougam.00,pothier.00}. 

\begin{figure}[ht]
\centerline{
\epsfxsize=10.0cm
\epsfbox{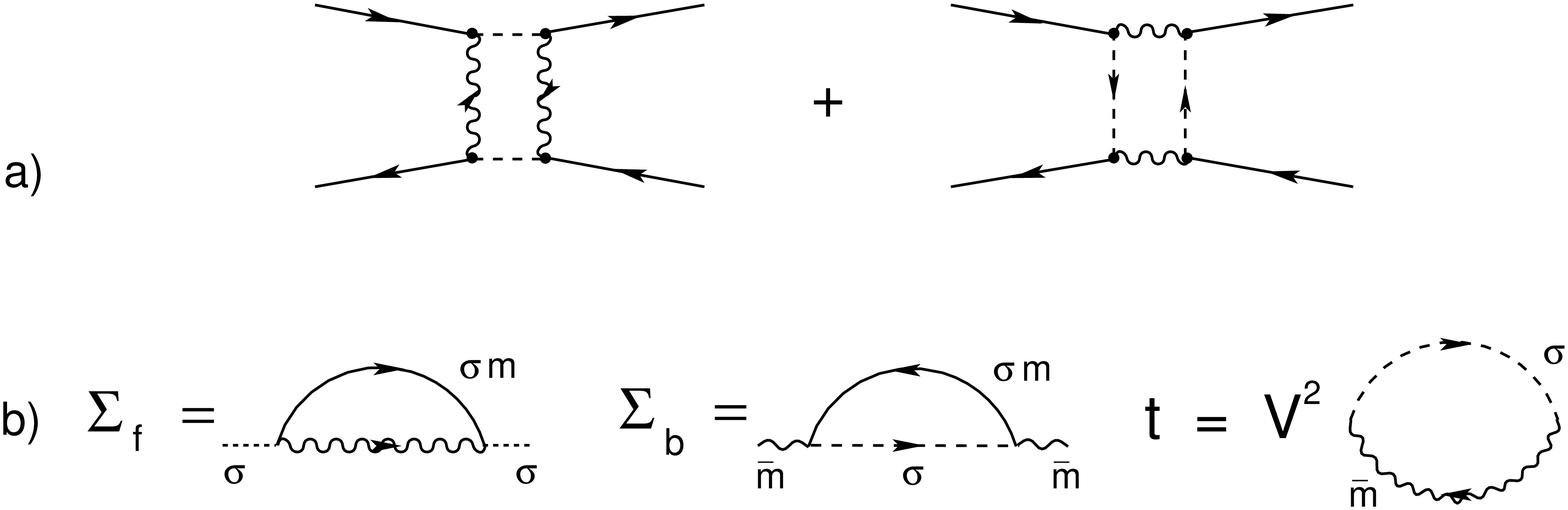}
}
\caption{a) Leading contributions to  the effective electron-electron
vertex induced by a Kondo impurity.  
b) Diagrammatic representation of the NCA auxiliary particle selfenergies, 
Eqs.~(\ref{NCA1})-(\ref{tmatrix}). Solid, dashed, and wiggly lines
correspond to conduction electron, pseudofermion, and slave boson
propagators, respectively (see text).}
\label{fig1}
\end{figure}

The infrared singular behavior of $\tilde V(\varepsilon)$, phenomenologically
deduced above, means in particular that the interaction has no 
essential momentum dependence and should be of a local origin, 
in contrast to an interaction mediated by a diffusive or dispersive 
collective mode. Therefore, it has been proposed \cite{kroha.00}
that the anomalous energy relaxation may be induced by Kondo impurities,
in particular by (nearly) degenerate atomic two-level
systems \cite{kroha.00} 
or other dynamical defects in the nanowire which can generate
a two-channel Kondo (2CK) effect \cite{nozieres.80,coxzawa.98}. 
The 2CK effect is known to have ideally a non-vanishing zeropoint
entropy $S(0) = k_B {\rm ln}\sqrt{2}$ \cite{wiegmann.83,andrei.84} 
and, hence, can cause dephasing at low energies \cite{zawa.99}. 
In fact, the electron interaction vertex mediated by a Kondo impurity
does show a $1/\varepsilon $ energy dependence. This was noted 
for the 2CK effect in the strong coupling region \cite{kroha.00}
(Fig.~\ref{fig1} a)) and independently
for the weak coupling (i.e. high energy) regime of the
magnetic, single-channel Kondo (1CK) effect \cite{glazman.00},
see also \cite{zawa.69}.
Although this analysis takes elevated single-particle energies into account,
it assumes the Kondo defect to be in thermodynamic equilibrium.  
Since, however, the effective interaction is generated dynamically
due to a Fermi edge singularity, it will be substantially modified in the
non-equilibrium situation of the energy relaxation measurements
considered here. 

Although the Kondo effect in stationary non-equilibrium
has been treated by various numerical methods before, no complete 
understanding has been reached up to now. In particular, the
scaling properties and to what extent the Kondo effect at finite 
bias is a strong coupling problem have remained unclear.
In this article we, therefore, summarize a predominantly 
analytical analysis of the 1CK and the 2CK 
non-equilibrium Kondo effect \cite{kroha.01}. It is shown that 
in nanowires with either 1CK or 2CK defects the quasiparticle distribution 
function exhibits scaling in terms of the applied bias, $eU$, when
$eU$ exceeds an intrinsic low-energy scale $eU_o$ which, up to logarithmic 
corrections, is proportional to the equilibrium Kondo temperature $T^{(0)}_K$.
The numerical solution of the quantum Boltzmann equation 
gives quantitative agreement with the experimental results discussed
above, where in the scaling regime the impurity concentration
is the only adjustable parameter. 
\inxx{captions,figure}

\section{A single Kondo impurity in a non-equilibrium wire}
The 1CK as well as the 2CK system can be described by a generalized
SU(N)$\times$SU(M) Anderson impurity model in the Kondo limit,  
i.e.~in terms of a quantum degree of freedom
in a low-lying level $\varepsilon_d$, coupled to the continuum of 
conduction electrons  
via a hybridization matrix element $V$,
where $N=2$ denotes the (pseudo)spin degeneracy and $M=1,2$ the number
of gegenerate, conserved conduction electron channels. 
In the auxiliary particle representation \cite{barnes.76} the model 
hamiltonian reads
\begin{eqnarray}
H=\sum _{\vec p,m,\sigma}\epsilon _p
c^{\dagger}_{pm\sigma}  c^{\phantom{\dagger}}_{pm\sigma}+
\epsilon _d \sum _{\sigma} 
f^{\dagger }_{\sigma } f^{\phantom{\dagger}}_{\sigma} 
+ V \sum _{p,m,\sigma} 
(f^{\dagger }_{\sigma} b_{\bar{m}} 
c^{\phantom{\dagger}}_{pm\sigma} + h.c.) \ ,
\label{eq:hamiltonian}
\end{eqnarray}
where $c^{\dagger}_{pm\sigma}$ is the creation operator for an electron
with momentum $\vec p$. The auxiliary fermion and boson
operators, $f^{\dagger }_{\sigma }$, 
$b^{\dagger }_{\bar m}$, create the local defect in
its quantum state $\sigma$ or in the unoccupied state, respectively.
The operator constraint 
$\hat Q = \sum _{\sigma} f^{\dagger }_{\sigma } f^{\phantom{\dagger}}_{\sigma}
        + \sum _{m}  b^{\dagger }_{\bar m} b^{\phantom{\dagger}}_{\bar m} 
\equiv 1$
ensures that the defect is in exactly one of its quantum states at any
instant of time. In the 1CK case of a magnetic Anderson 
impurity $\sigma $ denotes spin, and $m=1$ has no relevance.
For a 2CK defect, $\sigma$ is identified with 
a pseudospin, e.g. the parity of the local defect wave
function, and $m=1,2$ is the conduction electron spin
acting as the conserved channel degree of freedom. Note that in this case  
the bosonic operator $b^{\dagger }_{\bar m}$ transforms according to
the conjugate representation of SU(M) in order for the channel degree
of freedom to be conserved. The equilibrium Kondo temperature of the model 
is (to leading exponential order) $T^{(0)}_K \simeq W {\rm exp}(-1/NN_oJ)$,
with $J = V^2/|\varepsilon_d|$ the effective spin exchange coupling
and $W$ the half band width. 

The auxiliary particle representation allows for the application of 
standard Green's function methods 
and, in particular, is readily generalized \cite{hettler.94} 
to non-equilibrium by means of the Keldysh technique.
It is known \cite{cox.93} to describe the universal spectral 
properties of the 
multi-channel Kondo effect correctly both in the infrared
and in the high energy regime already within the non-crossing 
approximation (NCA), which is a selfconsistent, conserving approximation
to leading order in the effective  hybridization $\Gamma = V^2\pi N_o$. 
Thus, in the
multichannel case the NCA constitutes a correct infinite resummation 
of the logarithmic terms of perturbation theory. 
In the 1CK case {\it at finite bias} 
the logarithmic series is the same as for the 2CK effect {\it in 
equilibrium} because of the presence of two Fermi edges in the former
case \cite{kroha.01}. Thus, the NCA 
gives a correct description both of the 1CK and of the 2CK effects
in a strong non-equilibrium situation ($eU > T_K^{(0)}$) and will be used
for the explicit calculations below.    

In the remainder of this 
section we will consider a single Kondo defect at position
$x$ in a diffusive wire of length $L$ with a bias voltage $U$
applied between its ends.  
The generalization to the experimental situation of a dilute ensemble
of Kondo impurities will be treated in the following section.
The derivations apply both to the single and to the multi-channel case.
For purely elastic scattering in the wire 
the electronic distribution function at the position of the
impurity is 
\begin{eqnarray}
f_x(E,U)=\frac{x}{L} f^o(E+eU) +\Bigl(1- \frac{x}{L} \Bigr) f^o(E)\ .
\end{eqnarray}
Denoting the momentum integrated greater ($+$) and lesser ($-$) 
Keldysh Green's functions for the conduction electrons at position $x$ by
$G_{c,x}^{+}(E) \equiv \sum_p G_{c,x}^{+}(E,p)=2\pi i\, f_x(E,U)$ $N_o(E)$, 
\quad $G_{c,x}^{-}(E) \equiv \sum_p G_{c,x}^{-}(E,p)=$
$- 2\pi i(1-f_x(E,U))\, N_o(E)$, and similarly
the ``greater'', ``lesser'' and retarded auxiliary particle propagators by
$G_{f,b}^{+,-,r}(\omega )$, the non-equilibrium NCA equations then read 
(Fig.~\ref{fig1} b))
\begin{eqnarray}
\frac{G^{\pm}_f(\omega)}{|G^{r}_{f}(\omega)|^{2}} &=& 
   - M \Gamma  \int \frac{d\epsilon}{2\pi i}\, 
     G^{\pm}_{c,x}\, (- \epsilon ) G^{\pm}_b(\omega + \epsilon) 
\label{NCA1} \\
\frac{G^{\pm}_b(\omega)}{|G^{r}_{b}(\omega)|^{2}} &=& 
   + N \Gamma  \int \frac{d\epsilon}{2\pi i}\, 
     G^{\mp}_{c,x}\, (  \epsilon ) G^{\pm}_f(\omega + \epsilon) \ .
\label{NCA2}
\end{eqnarray}
This set of non-linear equations is closed by the Kramers-Kroenig relations, 
$G_{f,b}^{r}(\omega ) =\int d\varepsilon/(2\pi i)\, G_{f,b}^{-}(\omega )/
(\omega-\varepsilon+i0)$, which follow from analyticity and the fact
that the auxiliary particle Green's functions
have only forward in time propagating parts.
The conduction electron t-matrix due to the Kondo impurity is
\begin{eqnarray}
t_{c,x}^{\pm}(E ) =  
   - \frac{\Gamma}{\pi N_o(0)} \int \frac{d\epsilon}{2\pi i}\, 
     G^{\pm}_{f}\, (E +  \epsilon ) G^{\mp}_b(\epsilon) \ . 
\label{tmatrix}
\end{eqnarray}

It is crucial 
to determine the low-energy scales of the problem in non-equilibrium.
These may be determined from perturbation theory. 
At bias $eU$ the Kondo
temperature, definied as the breakdown scale of perturbation theory, is
suppressed, e.g. in the middle of the wire ($x=1/2$), as
\begin{eqnarray}
T_K (eU)= \sqrt{T_{K}^{(0)\ 2} + \Bigl(\frac{eU}{2k_B}\Bigr)^2}-\frac{eU}{2k_B}
\quad \stackrel{eU\gg k_BT^{(o)}_{K}} {\simeq} 
\quad \frac{k_BT_{K}^{(0)\ 2}}{eU} 
\label{eq:TK_eU}
\end{eqnarray}
In addition, a Korringa-like spin relaxation rate 
$1/\tau _s \propto eU$ appears,
because in non-equilibrium there is finite phase space available for
scattering even at $T=0$,
\begin{eqnarray}
\frac{1}{\tau_s(eU)} = 
\frac{1}{4\pi}\ NM\ (N_oJ)^2 \ eU  < eU\ .
\label{eq:tau_eU}
\end{eqnarray}
Because of this inelastic relaxation rate, the logarithmic singularities
of perturbation theory ($T=0$) are shifted by $1/\tau_s$ to the 
complex frequency plane, and the Kondo scale disappears eventually for large
bias $eU$ and is replaced by the inelastic rate
$1/\tau _s$: The low-temperature scale is $T_o(eU)={\rm max}
[T_K(eU),1/\tau _s(eU)]$.
The crossover from the Kondo to the Korringa scale occurs at a bias
\begin{eqnarray}
eU_o =   \frac {2}{N_o J}\sqrt{\frac{\pi}{NM}}\; k_BT_K^{(0)} =
         2N\sqrt{\frac{\pi}{NM}}\;
         k_BT_K^{(0)}\; {\rm ln}\Bigl( \frac{W}{T^{(0)}_K} \Bigr) \ .
\label{eq:Ucross}
\end{eqnarray}
This means that in the large bias regime, $eU>eU_{o}$, the low-energy
scale of the problem is proportional to the external bias itself.

We now turn to the scaling analysis of the non-equilibrium Kondo
equations (\ref{NCA1})--(\ref{tmatrix}).
It is well known \cite{muha.84,kroha.97}, that at $T=0$ in equilibrium
the exact auxiliary particle propagators exhibit
infrared powerlaw divergencies $G^{-}_{f,b}(\omega ) \propto  
\Theta (\omega )\ \omega ^{-\alpha_{f,b}}$, where the exponents
$\alpha _{f,b}$ are due to an orthogonality catastrophy between
the occupied and the unoccupied impurity state.
In the 2CK case the exact exponents are reproduced by NCA \cite{cox.93}
and depend in a characteristic way 
on the spin degeneracy $N$ and the channel number $M$, 
$\alpha _f = M/(N+M)$, $\alpha _b = N/(N+M)$ with $ \alpha _f+\alpha _b =1$.
In the large bias regime ($eU > eU_o$), 
the NCA equations are valid both in the 1CK and in the 2CK case, 
as mentioned above. The behavior around the Fermi edges may
be determined by defining the complex frequency variables
$z=\omega - i/2\tau_s(eU)$ etc.~and rewriting the NCA
equations for $z\to 0$ and $z\to eU$ 
as coupled differential equations in analogy 
to the equilibrium case \cite{muha.84}. We obtain damped powerlaw
behavior of the auxiliary fermion propagator for energies 
$|\omega | \stackrel{<}{\sim} 1/\tau _s$ and of the auxiliary
boson propagator for energies around the two
Fermi edges $|\omega | \stackrel{<}{\sim} 1/\tau _s$, 
$|\omega -eU| \stackrel{<}{\sim} 1/\tau _s$ where the
damping constant is $1/\tau _s$. The exponents are modified
in the large bias regime to $\alpha _f' = 2M/(N+2M)$, 
$\alpha _b' = N/(N+2M)$, where the additional factor of 2
originates from the fact that there are two separated Fermi
edges. The form of $\alpha _f'$, $\alpha _b'$ 
is reminiscent of an effective doubling of the channel number $M$
induced by the two Fermi edges. This behavior is confirmed by
the numerical evaluation of the problem and is 
consistent with a recent perturbative renormalization
group analysis \cite{coleman.00} of the problem. The fact that
powerlaw behavior (although damped) 
instead of logarithmic behavior persists
even at large bias may be seen as a remnant of the strong
coupling behavior of the Kondo problem in non-equilibrium.
It is crucial for scaling in terms of $eU$ to occur:
Putting the powerlaw behavior of the auxiliary propagators
back into Eqs.~(\ref{NCA1})--(\ref{tmatrix}), using 
$\alpha _f' +\alpha _b' =1$ and the important fact that 
the low-energy scale $T_o$ and 
the damping rate of the powerlaw behavior themselves scale
with the external bias, $T_o\ 1/\tau _s \propto eU$, it follows
immediately that these equations depend only on dimensionless
energies $\omega /eU$, $E/eU$ etc., i.e. they are scale invariant
in the regime $eU > eU_o =  2N\sqrt{\frac{\pi}{NM}}\;
k_BT_K^{(0)}\; {\rm ln}\Bigl( \frac{W}{T^{(0)}_K} \Bigr)$
(Eq.~\ref{eq:Ucross}).
It is seen that in experiments
one may extract the equilibrium Kondo temperature $T_K^{(0)}$
from the lower breakdown scale of the scaling behavior.

\section{Solving the kinetic equation}

\begin{figure}[ht]
\centerline{
{\epsfxsize=\linewidth
\epsfbox{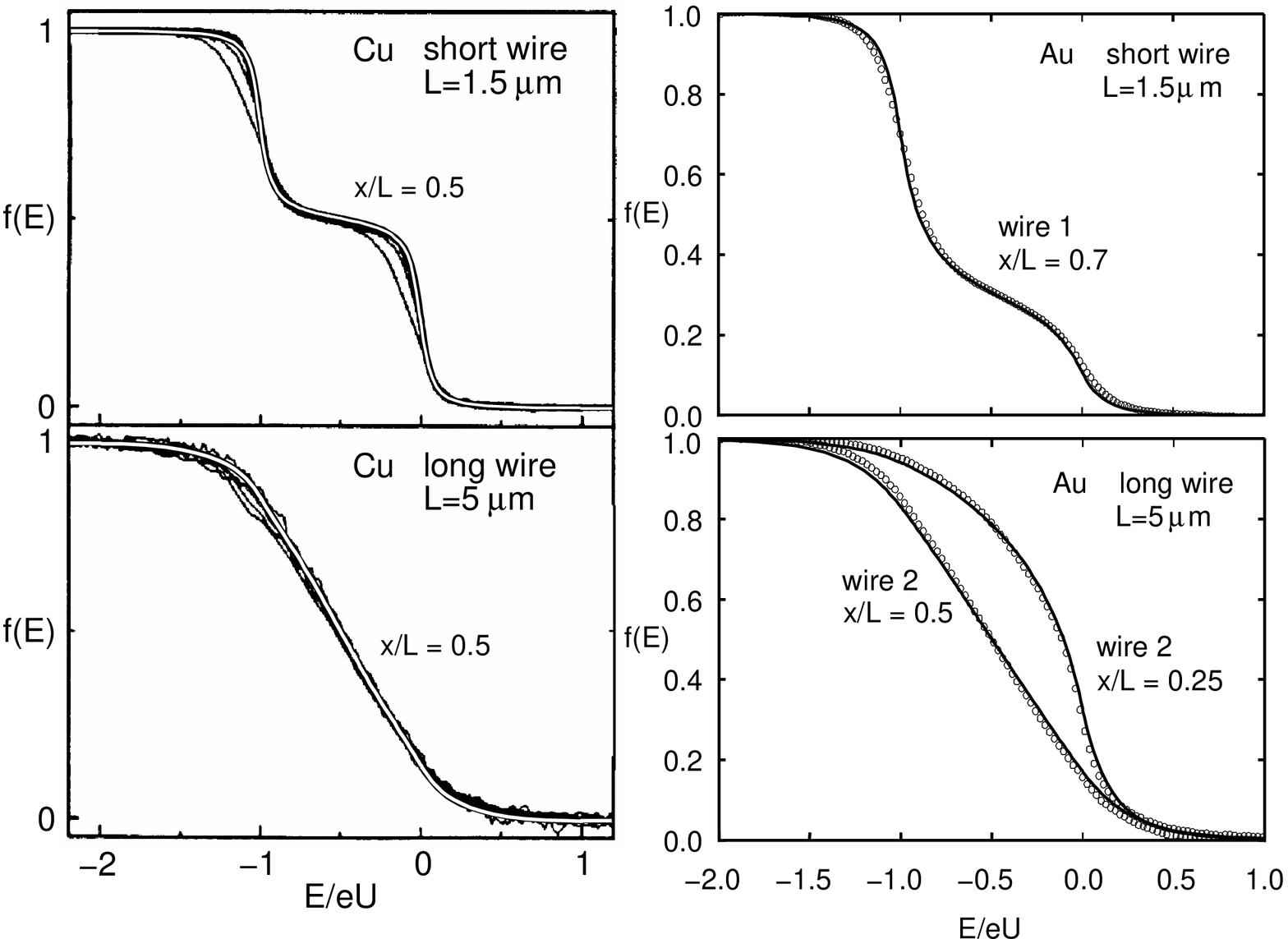}}
}
\caption{Scaled distribution functions $f_x(E,U)$ in nanowires. ---
Left panel: Two Cu nanowires
($L=1.5\mu m$, $D=65cm^2/s$, $x/L=0.5$ and 
$L=5.0\mu m$, $D=45cm^2/s$, $x/L=0.5$).
Black curves: Experimental data for 0.05mV$\leq U\leq$0.3mV 
in steps of 0.05mV \cite{pothier.97}. 
The two curves showing deviations from
scaling are at small voltages.
Light curves: Theory for 2CK impurities in the scaling regime, $eU > eU_o$.
Fitted 2CK impurity concentration: $c_{imp} \approx 
8\cdot 10 ^{-6}$/(lattice unit cell). --- 
Right panel: Two Au nanowires
(wire 1 and wire 2 of Ref.~\cite{pothier.00}),
measured at different positions $x/L$. Open circles: Experimental data;
Solid lines: Theory for magnetic impurities in the scaling regime.
Fitted magnetic impurity concentration: $c_{imp} \approx
2\cdot 10 ^{-4}$/(lattice unit cell).
} 
\label{fig2}
\end{figure}
\inxx{captions,figure}

The remaining task is to calculate the electronic
non-equilibrium distribution function $f_x(E,U)$ explicitly
at an arbitrary position $x$ along the wire 
in the presence of a dilute but finite density of 1CK or 2CK defects
$c_{imp}$.

The Boltzmann quantum kinetic equation for the momentum $\vec p$
dependent ``greater'' conduction electron Green's function 
$G_{c, (x,t)}^{+} (E,\vec p)$, 
which contains information about the distribution function, reads
\cite{landau.97}
\begin{eqnarray}
\Bigl[ \frac{\partial }{\partial t} +
\nabla _x \cdot \frac{\vec p}{m}
+ e\vec E (\vec x) \cdot \nabla _p\Bigr]
G_{c, (x,t)}^{+} (E,\vec p) =  {\cal C}
\bigl( \{ G_{c,(x,t)}^{+}(E,\vec p) \} , c_{imp} \bigr) , \nonumber\\
\label{boltz} 
\end{eqnarray}
where ${\cal C}$ is the collision integral induced by inelastic 
scattering off Kondo defects and
$\vec E(\vec x)$ are the external electric fields, including
those generated by static, random impurities in the wire.
Exploiting the fact that the current in the disordered system is diffusive,
\begin{equation}
\vec j _x(E) =\sum _p \frac{\vec p}{m}  
G_{c, x}^{+} (E,\vec p)  = - D \nabla _x 
\rho _x(E) = - D \nabla _x  \sum _p 
G_{c, x}^{+} (E,\vec p) \ ,
\end{equation}
with $D$ the diffusion coefficient,
and summing in Eq.~(\ref{boltz}) over quasiparticle momenta $\vec p$,
the electric field term drops out as a surface term, and we obtain
in the stationary case a diffusive kinetic equation \cite{nagaev.92}
for the impurity averaged distribution function 
$\overline{f}_x(E,U)= \sum _p G_{c, x}^{+} (E,\vec p)/$ $2\pi i N_o$
as function of the quasiparticle energy $E$ and position $x$, 
\begin{equation}
-D \nabla _x^2 \overline{f}_x(E,U) =  {\cal C}
\bigl( \{ \overline{f}_x(E,U) \} , c_{imp} \bigr) \ .
\label{boltzdiff}
\end{equation}
After momentum integration, the collision integral is given in terms of the 
impurity t-matrix (Eq.~\ref{tmatrix}) to arbitrary order in the
electron-impurity coupling $\Gamma$ by
\begin{equation}
{\cal C}(\{ \bar f _x(E,U) \}, c_{imp} ) = \frac{c_{imp}}{2\pi N_o} 
[ t_{c,x}^-(E) G_{c,x}^+(E) - t_{c,x}^+(E) G_{c,x}^-(E) ] \ .
\label{eq:collint} 
\end{equation}
We have solved the set of non-linear integro-differential
equations (\ref{NCA1})--(\ref{tmatrix}), (\ref{boltzdiff}),
(\ref{eq:collint}) numerically with the boundary conditions 
$\overline{f}_{x=0}(E,U)$ $ = f^o(E)$,
$\overline{f}_{x=L}(E,U) = f^o(E+eU)$ and that the electron system should be
in local equilibrium with the dynamical impurity.
The results are shown in Fig.~\ref{fig2}.
The theoretical $\overline{f}_{x}(E/eU)$
curves show scaling behavior in the regime $eU>eU_o$ (Eq.~(\ref{eq:Ucross}))
for bias voltages $U$ varying within a factor of 4 to 9, 
depending on the high energy parameters of the model Eq.~(\ref{eq:hamiltonian}).
There is excellent
quantitative agreement between theory and experiment for all samples.
The experimental data for Cu and Au wires
can be fitted equally well by the 1CK or by the 2CK theory, with 
very similar defect concentrations in the two cases. As an example,
Fig.~\ref{fig2} shows the fits of the Cu data to the 2CK theory and of the
Au data to the 1CK theory only. 
The agreement is consistent with the conjecture that the 
anomalous energy relaxation in those wires is mediated either 
by 1CK or by 2CK defects .

\section{Relation to dephasing}
For an ideal 2CK system in equilibrium, the non-vanishing quasiparticle 
scattering rate $1/\tau (E)$ crosses over to a pure dephasing 
rate $1/\tau _{\varphi}$,
as the quasiparticle energy approaches the Fermi energy, $E,T\to 0$
\cite{zawa.99}. However, because of the finite level splitting $\Delta$
in real systems, one expects an upturn of $\tau _{\varphi}$
at the lowest $T\stackrel {<}{\sim} \Delta /k_B$, with an intermediate
plateau for $\Delta /k_B \stackrel {<}{\sim} T \stackrel {<}{\sim} T_K^{(0)}$
\cite{zawa.99}. 
Such behavior can also be expected from magnetic (1CK) Kondo impurities 
in combination with a phonon contribution, because the spin
flip rate reaches a maximum at $T\simeq T_K^{(0)}$ \cite{muha.71}. 
This has been observed in the Kondo system AuFe \cite{bergmann.87}  
and most recently in clean Au wires presumably containing a small 
concentration of Fe impurities \cite{pothier.00}.
One might, therefore, conjecture that 1CK or 2CK defects could be the
origin of the dephasing time saturation observed in magnetotransport
measurements of weak localization \cite{mohanty.97,gougam.00,pothier.00}. 
This assumption is indeed supported by
several coincidences between the dephasing time measurements and the  
results on the non-equilibrium distribution function: (1) The
dephasing time $\tau_{\varphi}$ extracted from magnetotransport
experiments \cite{gougam.00,mohanty.97} 
is strongly material, sample, and preparation dependent.
This suggests a non-universal dephasing mechanism, like dynamical
defects, which is not inherent to the electron gas.
(2) The dephasing time in Au wires is generically shorter 
than in Cu wires \cite{gougam.00}.
This is consistent with the fact that the estimates
for the dynamical impurity concentration $c_{imp}$, obtained from the
fit of the present theory to the experimental distribution functions,
is substantially higher in Au than in Cu wires (Fig.~\ref{fig2}).
(3) In Ag wires one observes neither dephasing saturation 
nor $E/eU$ scaling of the distribution function \cite{pierre.00}. 
This is consistent with the assumption that dephasing saturation and
anomalous energy relaxation in the nanoscopic wires have the same
origin and that there are no Kondo defects present in the Ag samples.

\section{Concluding remarks}

We have analysed the single- as well as the two-channel Kondo effect 
in a stationary non-equilibrium situation.
It was found that a Korringa-like, inelastic spin relaxation rate
appears which at large bias sets the low-energy scale of the problem. 
Nevertheless, a remnant of the strong coupling Kondo fixed point
persist even at large bias, which manifests itself by damped
powerlaw behavior of the local spectral density. The latter leads
to scaling of the non-equilibrium distribution function
$f_x(E,U)=f_x(E/eU)$ as an experimentally observable signature.
The present theory yields quantitative agreement with the
experimental results for all samples measured, the density of 
Kondo defects in the wire being the only 
adjustable parameter in the scaling regime.
This strongly suggests that the anomalous energy relaxation
might be caused by either 1CK or 2CK defects. However, it was shown
that the scaling property of the distribution function does not
distinguish between 1CK or 2CK impurities. Further experimental tests,
like application of a magnetic field, will be required for that 
purpose.

\section*{Acknowledgements}
It is a pleasure to thank A. Zawadowski, H. Pothier, B. L. Al'tshuler,
D. Esteve, J. v. Delft, and P. W\"olfle for stimulating and fruitful discussions.
The members of the 
Saclay group have provided experimental data prior to publication
which is gratefully acknowledged. 
This work is supported by DFG through SFB 195.

\end{document}